# The effect of spin drift on spin accumulation voltages in highly-doped Si

Makoto Kameno [1], Eiji Shikoh [1], Teruya Shinjo [1], Tomoyuki Sasaki [2], Tohru Oikawa [2], Yoshishige Suzuki [1], Toshio Suzuki [3] and Masashi Shiraishi [1]

1. Graduate School of Engineering Science, Osaka Univ. Toyonaka, Japan
2. SQ Research Center, TDK Corporation, Nagano, Japan
3. AIT, Akita Research Institute of Advanced Technology, Akita, Japan

**Abstract**

An investigation was carried out into the effect of spin drift on spin accumulation signals in highly-doped Si using non-local 4-terminal (NL-4T) and 3-terminal (NL-3T) methods. The spin signals in the NL-4T scheme were not affected by spin drift, and the bias dependence was governed by whether spins were injected into or extracted from the Si channel. In contrast, the spin signal was strongly modulated by the bias electric field in the NL-3T scheme. The bias electric field dependence of the spin signals in the NL-3T method was quantitatively clarified using the spin drift-diffusion equation, and the results can be reasonably explained.

Corresponding author: M. Shiraishi (shiraishi@ee.es.osaka-u.ac.jp)

Spin drift, which is usually negligible in spin transport in metallic systems, contributes significantly to spin transport and spin accumulation voltages in semiconductors. Since an electric field gives rise to spin drift, investigating spin accumulation voltages as a function of the bias electric field (bias voltage) in semiconductor spin devices can clarify how spin drift governs spin transport and accumulation properties. Thus, studying the bias dependence of the spin signal is an appropriate method for understanding the spin drift effect in semiconductors. Although there have been a number of studies on Si spintronics, achieving spin transport [1-5] and spin accumulation [6-10], the bias dependence of the spin signals induced by spin drift in Si remains inconclusive. This bias dependence is currently one of the most important research issues in the basic physics of spin transport in Si. In previous studies, the dependence was investigated using a non-local 3-terminal (NL-3T) geometry [6-10] in which the accumulated spins beneath a ferromagnetic (FM) electrode were affected by an applied electric field. However, it is quite remarkable that the bias dependence in NL-3T did not exhibit obvious differences under positive and negative bias conditions, although spin drift governs the spin accumulation voltage, unlike in the non-local 4-terminal (NL-4T) geometry, and is also remarkable that the reported behavior of the bias dependence did not have accordance each other [6-10]. Furthermore, the samples used in the previous studies did not exhibit spin transport at room temperature, and the correspondence between spin accumulation and spin transport was not fully clarified. Therefore, the nature of spin drift in Si remains an open question, and no precise understanding has been established. The present paper reports on careful and systematic studies of spin accumulation voltages as a function of the bias voltage in a Si spin valve, where spin injection and spin transport were simultaneously realized at up to 294 K [1]. The bias voltage dependence of the spin voltages in NL-3T and NL-4T exhibited obviously different behavior, and this finding can be quantitatively understood using a spin drift-diffusion model.

The present device consisted of a P-doped ($\sim 5\times 10^{19}$ cm$^{-3}$) n-type Si channel with a width and thickness of 21 μm and 80 nm, respectively, on a SOI substrate. The device was equipped with two ferromagnetic electrodes (FM1 and FM2) and two nonmagnetic electrodes (NM1 and NM2). The FM and NM electrodes were made of Fe on an MgO tunneling barrier layer and Al, respectively. The widths of FM1 and FM2 were 0.45 μm and 2.0 μm, respectively, and the center-to-center distance between the FM electrodes was 1.72 μm. The device fabrication method was described in detail elsewhere [2]. The conductivity and diffusion constant of the Si channel were measured to be $1.03\times 10^{5}$ Ω$^{-1}$ m$^{-1}$ and 4.32 cm$^2$/s at 8 K, respectively. NL-3T and NL-4T methods for detecting Hanle-type spin precession were employed under dc electric current injection (see Figs. 1(a) and (b)). The measurements were repeated several times, and the data was averaged to obtain good fits to theoretical curves. The spin diffusion length λ, spin lifetime τ, and spin polarization P were estimated using the following fitting function,

$$\frac{V}{I} = \frac{P^2}{\sigma A / D}\int_0^\infty \frac{1}{\sqrt{4\pi Dt}}\exp\left(-\frac{L^2}{4Dt}\right)\cos(\omega t)\exp\left(-\frac{t}{\tau}\right)dt \ , \quad (1)$$

but the analytical solution obtained by taking the width of the FM electrodes into account was used for the fitting [1, 11]. The gap length between the two FM electrodes was set to zero in the solution for the NL-3T case. The bias dependence of the spin voltages was investigated by applying a dc current from -4.0 mA to 4.5 mA, where a positive (negative) bias current was defined as the case when spins were injected into (extracted from) the Si channel. Before conducting the Hanle measurements, the non-local magnetoresistance of the sample was observed up to 294 K (see Figs. 1(c) and (d)) to confirm that spins were injected and transported into the Si channel. This was important, because previous studies focused only on spin accumulation voltages in NL-3T. Note that all of the measurements of the bias dependence in this study were implemented at 8 K, because the resulting good signal/noise ratio in the spin signals allowed precise and reliable estimation of the spin transport parameters.

Figure 2(a) shows NL-4T Hanle signals at 8 K when the dc bias currents were changed from -4.0 to +4.5 mA. The results for the intensities of the spin signals and the $P$, $\lambda$ and $\tau$ values are shown in Figs. 2(b)-(e). The NL-4T signals monotonically decreased when higher bias currents were applied, which can be attributed to a reduction in $P$ for the spins injected into the Si because $\lambda$ (~1.8 μm) and $\tau$ (~8 ns) were almost constant. The reduction of $P$ has been thought to be due to phonon and/or magnon scattering at the FM/MgO interface [12]. It is notable that the spin signals under a negative bias current were smaller than those under the positive bias currents. Similar behavior has been reported [3,13,14], and it is generally explained as follows; in a simple model assuming a parabolic band structure in FM [14], the spin polarization of FM above the Fermi level becomes lower than at the Fermi level, which can explain the stronger bias dependence of spin signals under a negative bias current (the spin extraction case) in our study.

Unprecedented features were observed in the NL-3T measurements, in which spin drift governed the spin accumulation voltages. Figure 3(a) shows the spin voltage in the NL-3T Hanle experiments under -1.0 and +1.0 mA bias currents. Here, the spin accumulation voltage was observed only under a negative bias current (-1.0 mA), and not under positive current application up to +4 mA. The bias dependence of the NL-3T Hanle signals and the estimated spin diffusion length are shown in Figs. 3(b) and (c). Although the NL-3T Hanle signals as a function of bias current behaved similarly to those in the NL-4T measurements, a decrease in spin lifetime as a function of bias current was observed. More importantly, the spin lifetime values were smaller than those in the NL-4T measurements. This was a result of a spin-drift-based mechanism, as will be discussed later.

Although the doping concentration of the Si channel was above the effective density of states in the conduction band, spin drift cannot be completely neglected under an applied electric field. The spin accumulation signal in the NL-4T method is not affected by spin drift because no electric field is applied in the detecting circuit, and the driving force for the spins in the detection circuit is spin diffusion. Hence, the intensity of the output spin voltages is governed by the number of accumulated spins and the spin diffusion length. In contrast, the accumulated spins beneath the FM electrode in the NL-3T scheme is affected by the electric field applied for bias current injection. This difference in the measurements gives rise to a modulation of the spin voltages, as was the case in our study. Physically, the application of the electric field modulated the spin diffusion length, resulting in a modulation of the spin lifetime. The spin drift-diffusion equation in a Fermi-liquid system can be described as,

$$\nabla^2(n_\uparrow - n_\downarrow) - \frac{\mu E}{D} \cdot \nabla(n_\uparrow - n_\downarrow) - \frac{1}{\lambda^2}(n_\uparrow - n_\downarrow) = 0, \qquad (2)$$

where $n_\uparrow - n_\downarrow$ is the difference between the up-spin and down-spin electron densities, $\mu$ is the mobility of electrons in the channel, $D$ is the diffusion constant of the channel, and $E$ is the applied electric field [15]. Due to the existence of an electric field, the spin diffusion length is enhanced (suppressed) by the electric field along (against) the direction of spin motion, as discussed above. The general form of the spin diffusion length under a negative bias current $\lambda_u$ can be expressed using the Einstein relation as,

$$\lambda_u^{-1} = \frac{|3eE|}{4E_F} + \sqrt{\left(\frac{|3eE|}{4E_F}\right)^2 + \frac{1}{\lambda^2}}, \qquad (3)$$

where $E_F$ is the Fermi energy of the spin channel (the highly-doped Si). Figure 3(d) shows the spin diffusion length, which was experimentally obtained using the NL-3T method, as a function of the bias voltage. Theoretical fitting using Eq. (3) can explain the general trend of the spin diffusion length well (see Fig. 3(d)), and enables an estimation of the Fermi energy of the highly-doped Si to be 0.08 eV. The Fermi energy can be analytically calculated to be 0.05 eV using the Einstein relation and the experimentally obtained $\mu$ and $D$ values of the sample, and it is notable that these experimental and theoretical estimations are in good accordance. Therefore, the characteristic bias dependence of the NL-3T spin accumulation voltages in the highly-doped Si channel can be explained by a spin drift-diffusion model. The disappearance of the spin signals under positive bias can be qualitatively explained by a reduction of the spin diffusion length, although a quantitative explanation has not been fully established. The spin sensitivity at the FM interface as a function of the bias electric field should be taken into account for further quantitative understanding [16], since the sensitivity has an $E^{-2}$-dependence that strongly suppresses the spin signals, as has been observed in GaAs [17].

In summary, systematic studies were carried out into spin accumulation voltages in NL-4T and NL-3T. It was found that spin drift played an important role in the bias dependence of the spin signals in NL-3T. The bias electric field dependence of the spin signals was quantitatively investigated, and it was found that the bias dependence was reasonably explained by a spin drift-diffusion model. Whilst previous studies did not successfully explain the nature of spin transport in Si, the present study provides a good foundation for future discussion of the spin transport in Si.

**References**

1. T. Suzuki, T. Sasaki, T. Oikawa, M. Shiraishi, Y. Suzuki, and K. Noguchi, Appl. Phys. Express. **4**, 023003 (2011).

2. T. Sasaki, T. Oikawa, T. Suzuki, M. Shiraishi, Y. Suzuki, and K. Noguchi, Appl. Phys. Lett. **96**, 122101 (2010).

3. M. Shiraishi, Y. Honda, E. Shikoh, Y. Suzuki, T. Shinjo, T. Sasaki, T. Oikawa and T. Suzuki, Phys. Rev. B **83**, 241204(R) (2011).

4. I. Appelbaum, B. Huang, and D. J. Monsma, Nature **477**, 17 (2007).

5. B. Huang, D. J. Monsma, and I. Appelbaum, Phys. Rev. Lett. **99**, 177209 (2007).

6. C.H. Li, O.M.J. van 't Erve, and B.T. Jonker: Nature Comm. **2,** 245 (2011).

7. S. P. Dash, S. Sharma, R. S. Patel, M. P. de Jong and R. Jansen: Nature **462**, 26 (2009).

8. K-R. Jeon, B-C. Min, I-J. Shin, C-Y. Park, H-S. Lee, Y-H. Jo, and S-C. Shin, Appl. Phys. Lett. **98**, 262102 (2011).

9. Y. Ando, K. Kasahara, K. Yamane, Y. Baba, Y. Maeda, Y. Hoshi, K. Sawano, M. Miyao and K. Hayama, Appl. Phys. Lett. **99**, 012113 (2011).

10. N.W. Gray and A. Tiwari, Appl. Phys. Lett. **98**, 102112 (2011).

11. H. Dery, L. Cywinsly and L.J. Shan, Phys. Rev. B **73**, 41306 (2008).

12. S. Yuasa, T. Nagahama, A. Fukushima, Y. Suzuki, and K. Ando Nat. Mater. **3**, 868 (2004).

13. M. Shiraishi, M. Ohishi, R. Nouchi, T. Nozaki, T. Shinjo and Y. Suzuki, Adv. Func. Mat. **19**, 3711 (2009).

14. S. O. Valenzuela, D. J. Monsma, C. M. Marcus, V. Narayanamurti and M. Tinkham, Phys. Rev. Lett. **94**, 196601 (2005).

15. Z.G. Yu and M.E. Flatte, Phys. Rev. B **66**, 201202(R) (2002).

16. A. N. Chantis and D. L. Smith, Phys. Rev. B **78**, 235317 (2008).

17. S. A. Crooker, E. S. Garlid, A. N. Chantis, D. L. Smith, K. S. M. Reddy, Q. O. Hu, T. Kondo, C. J. Palmstrom, and P. A. Crowell, Phys. Rev. B **80**, 041305(R) (2009).

**Figure Captions**

**Figure 1**

Typical sample structures and measurement geometries in (a) non-local 4-terminal, and (b) non-local 3-terminal. An external magnetic field was applied along the *z*- and *y*-axis for Hanle and non-local magnetoresistance measurements, respectively. Non-local magnetoresistance observed at (c) 8 K and (d) 294 K, where clear hysteresis was observed. The bias electric current was set to be +3.0 mA.

**Figure 2**

(a) Hanle-type spin precession signals in the NL-4T method at 8 K, where the bias electric currents were changed from -4.0 mA to +4.5 mA in 0.5 mA steps. The inset shows the Hanle-type spin precession signals under parallel (red) and anti-parallel (blue) magnetization configurations. The solid lines are theoretical fitting lines.

(b)-(d) The bias current dependence of (b) the spin accumulation voltages, (c) the spin polarization, (d) the spin diffusion length and (e) the spin relaxation time. An asymmetric dependence on the bias current was observed in the spin voltages and spin polarization, while the spin diffusion length and spin relaxation time were almost constant.

**Figure 3**

(a) Spin accumulation voltages in the non-local 3-terminal measurements under +1.0 mA (black open circles) and -1.0 mA (red open circles). The solid red line is the theoretical fitting line. Note that no signal was detected at 1.0 mA, even after repeating the measurements several times and averaging the data.

**(b)** Bias current dependence of the spin accumulation voltages under negative current conditions. Since no signal was detected under the application of a positive current, only the dependence under negative current conditions is shown.

**(c)** Spin relaxation time as a function of negative bias current. The signal monotonically decreased as the negative current increased.

**(d)** Spin diffusion length, which was calculated using the estimated spin diffusion length and the relation $\lambda = \sqrt{D\tau}$. The red dashed line is the least-squares fitting line using Eq. (3).

Fig.1

(a)

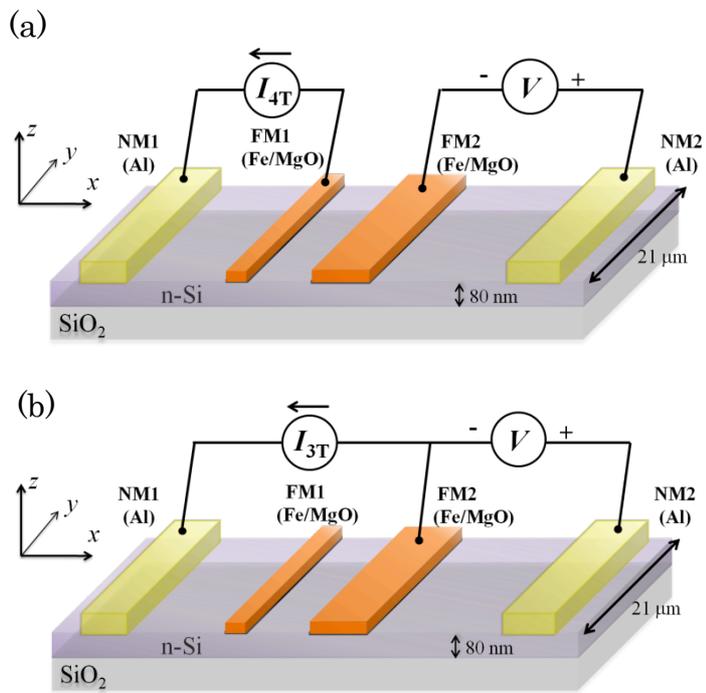

(b)

(c)

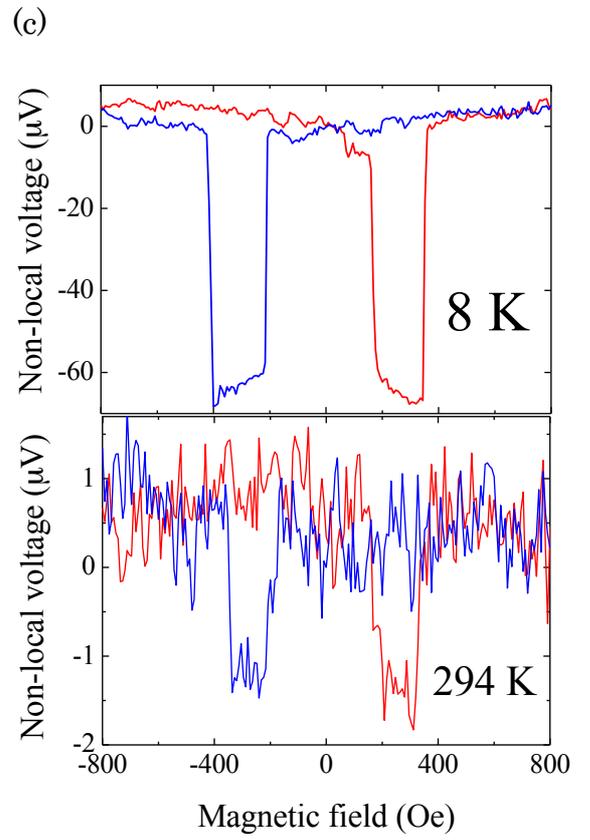

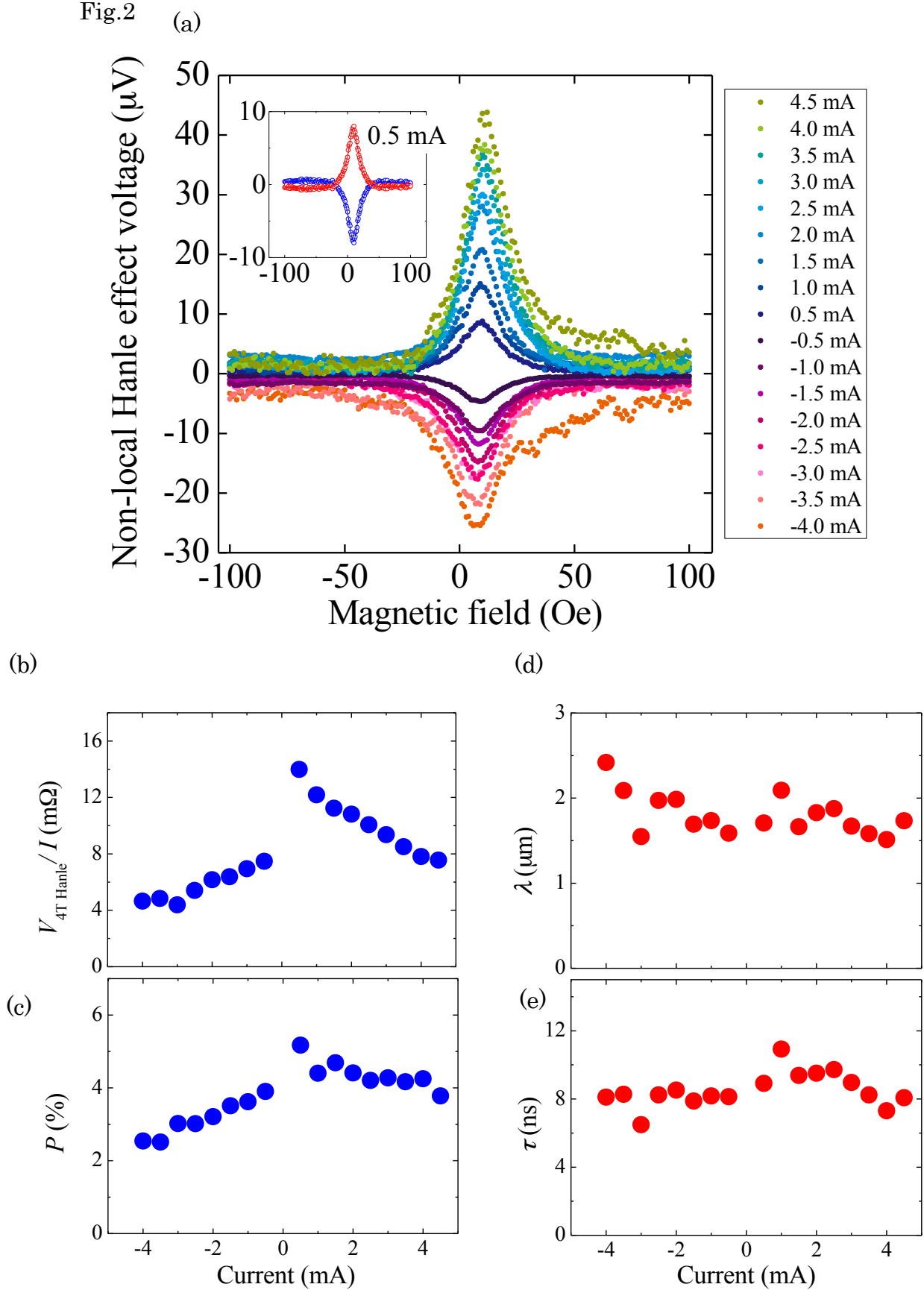

Fig.3

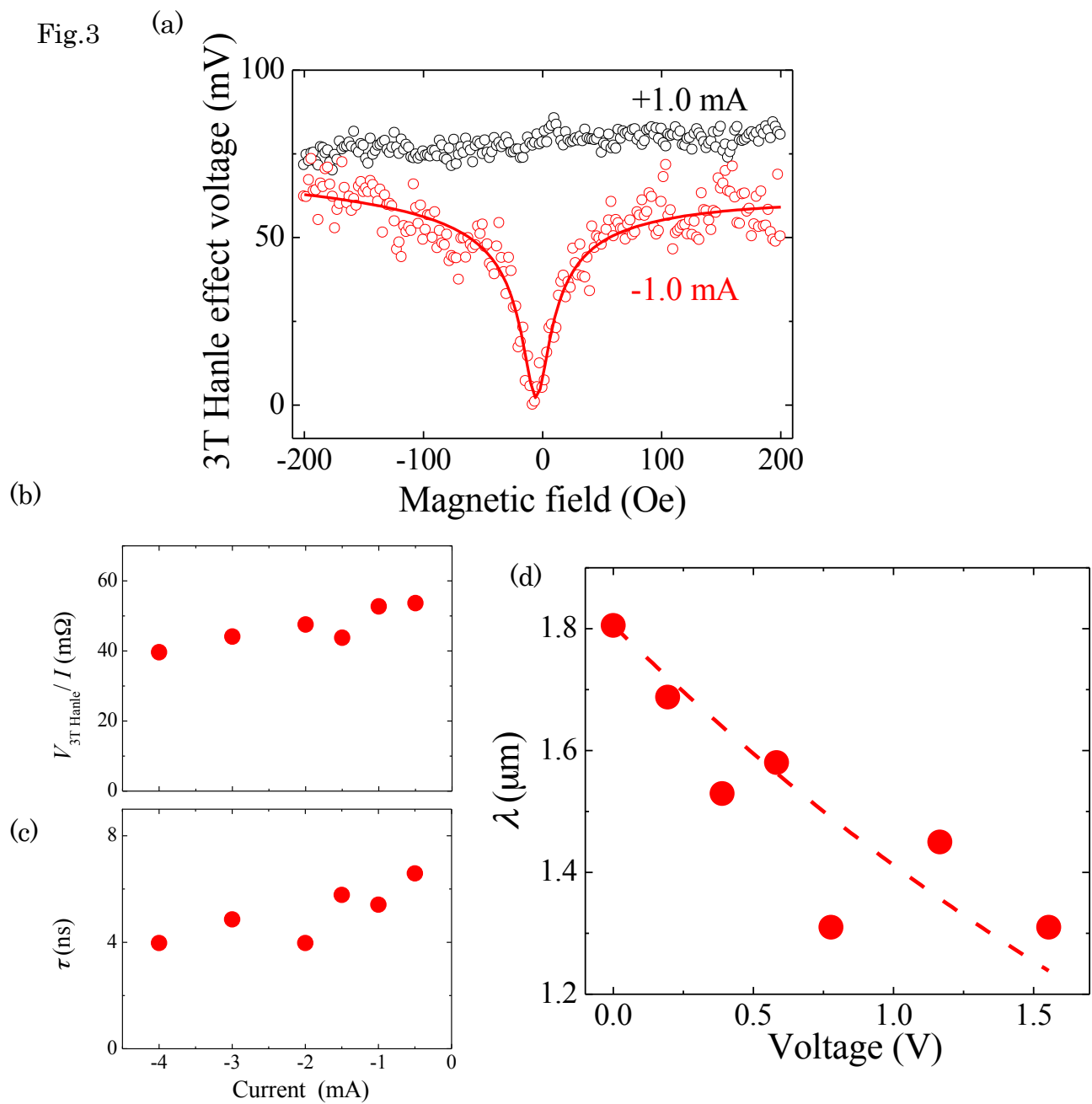